# Interpretable Anomaly Detection in Encrypted Traffic Using SHAP with Machine Learning Models

Kalindi Singh[1]       Aayush Kashyap[1]       Aswani Kumar Cherukuri[1*]

School of Computer Science Engineering and Information Systems, Vellore Institute of Technology, Vellore 632014, India

Email: cherukuri@acm.org

## Abstract

Purpose:

The widespread adoption of encrypted communication protocols such as HTTPS and TLS has enhanced data privacy but also rendered traditional anomaly detection techniques less effective, as they often rely on inspecting unencrypted payloads. This study aims to develop an interpretable machine learning-based framework for anomaly detection in encrypted network traffic.

Design/methodology/approach:

This study proposes a model-agnostic framework that integrates multiple machine learning classifiers — XGBoost, Random Forest, and Isolation Forest — with SHapley Additive exPlanations (SHAP) to ensure post-hoc model interpretability. The models are trained and evaluated on three benchmark encrypted traffic datasets: CIC-Darknet2020, USTC-TFC2016, and CSE-CIC-IDS2018. Performance is assessed using standard classification metrics, and SHAP is used to explain model predictions by attributing importance to individual input features.

Findings:

The XGBoost model achieved a peak classification accuracy of 99.94%, outperforming other models across multiple datasets. SHAP visualizations successfully revealed the most influential traffic features contributing to anomaly predictions, enhancing the transparency and trustworthiness of the models.

Originality:

Unlike conventional approaches that treat machine learning as a black box, this work combines robust classification techniques with explainability through SHAP, offering a novel interpretable anomaly detection system tailored for encrypted traffic environments.

Research Limitations & Implications:

This study is limited to three publicly available encrypted traffic datasets. While the framework is generalizable, real-time deployment and performance under adversarial conditions require further investigation. Future work may explore adaptive models and real-time interpretability in operational network environments.

Practical implications:

This interpretable anomaly detection framework can be integrated into modern security operations for encrypted environments, allowing analysts not only to detect anomalies with high precision but also to understand why a model made a particular decision — a crucial capability in compliance-driven and mission-critical settings.

## 1. Introduction

The exponential growth of encrypted network traffic, driven by the widespread adoption of protocols like HTTPS, TLS, and VPN tunnelling, has significantly bolstered data confidentiality and user privacy across the internet. A network attack is where an attacker gains unauthorised access to the network to perform malicious activities [4]. As organizations increasingly rely on encrypted channels to safeguard sensitive information, these protocols have become the default for secure communications. However, this surge in encryption has posed a unique set of challenges for cybersecurity analysts and network administrators. Traditional anomaly detection systems, which rely heavily on deep packet inspection (DPI) and payload analysis, struggle to operate effectively in encrypted environments, as they are unable to access the actual content of the data packets [1], [2].

Consequently, the focus of network security research has shifted toward metadata-driven detection strategies, relying on flow-based features such as packet length, inter-arrival times, and directionality. Although these features can be analysed using conventional statistical methods, the advent of machine learning (ML) has significantly improved anomaly detection performance by uncovering complex, non-linear patterns within network traffic data [3], [4]. Despite these advantages, ML models are often considered "black boxes" due to their lack of transparency, which is a critical limitation in high-stakes domains like cybersecurity where understanding the rationale behind a prediction is vital [5]. This growing tension between detection accuracy and model interpretability has given rise to the field of Explainable Artificial Intelligence (XAI), which seeks to make ML decisions more understandable to human stakeholders.

In traditional network anomaly detection, security mechanisms typically depend on the ability to inspect packet payloads for suspicious signatures or rule violations. However, with most modern internet traffic being encrypted, such payloads are no longer directly accessible, effectively rendering signature-based and rule-based systems less effective or even obsolete [1], [6]. While ML-based approaches can mine flow-level features to detect anomalies in encrypted environments, their lack of interpretability impedes trust and adoption in practical deployments. For example, network administrators and security analysts need to understand why an alert was triggered in order to respond appropriately, prioritize incidents, or even comply with regulatory mandates [5], [7].

Explainable AI (XAI) emerges as a viable solution to this problem by offering tools and frameworks that elucidate the inner workings of complex models. Techniques such as SHapley Additive exPlanations (SHAP) provide fine-grained, feature-level attributions that make it possible to identify the most influential aspects of a prediction [8]. By integrating XAI with ML-based anomaly detection systems, it becomes feasible to analyse encrypted network traffic in a way that is both accurate and interpretable.

This work aims to build an interpretable anomaly detection system for encrypted network traffic by integrating SHAP with individual ML models. Our goals are to (1) detect anomalies in encrypted traffic with high accuracy and (2) provide interpretable insights into the detection process. This project aims to investigate how Explainable AI techniques, particularly SHAP, can be effectively employed to detect and interpret anomalies in encrypted network traffic. The primary purpose is to bridge the gap between high-performance anomaly detection enabled by ML models: XGBoost, Random Forest, and Gradient Boosting, and the transparency required for practical cybersecurity applications.

The key contributions of this work are as follows:

- A comprehensive analysis of the limitations of traditional anomaly detection techniques in the context of encrypted network traffic, motivating the need for interpretable ML models.
- Implementation of multiple machine learning models—XGBoost, Random Forest, and Gradient Boosting—for anomaly detection using flow-based features from encrypted traffic.
- Application of SHAP for post-hoc interpretability, enabling an in-depth understanding of the contribution of each feature to the individual model's decisions.
- Empirical evaluation demonstrating that SHAP-based explanations can reveal actionable insights and improve the trustworthiness and usability of ML-driven anomaly detection systems.
- Discussion of the practical implications of combining XAI and ML in cybersecurity, with recommendations for deploying interpretable detection systems in operational environments.

By focusing on these contributions, this project underscores the potential of XAI to not only enhance the technical performance of anomaly detection systems but also to make them more transparent, trustworthy, and actionable in real-world scenarios.

## 2. Related Work

Traditional approaches to network anomaly detection include rule-based systems, signature matching, and statistical thresholding techniques. These methods often rely on Deep Packet

Inspection (DPI), which scrutinizes the packet payload for known patterns of malicious behaviour [1], [2]. Signature-based systems, such as Snort and Suricata, are effective against known threats but fail to generalize to new or obfuscated attacks [3]. Additionally, threshold-based statistical methods can detect volumetric anomalies but struggle with subtle deviations indicative of stealthy attacks [4]. The effectiveness of these methods is significantly reduced in encrypted environments, as encryption obfuscates the payload, rendering DPI-based detection infeasible.

To overcome limitations of traditional methods, machine learning has been increasingly adopted for anomaly detection in encrypted network traffic. ML models operate on flow-level features like packet size, inter-arrival time, and flow duration, which remain observable even when payloads are encrypted [3], [4], [9]. ML methods like XGBoost and Random Forest have shown considerable promise in detecting anomalies in flow metadata due to their ability to model complex, nonlinear patterns [6], [10]. For instance, Ikram et al. [4] employed an XGBoost ensemble integrated with deep neural networks to outperform baseline models in encrypted traffic classification, all without decrypting the data. These advances represent a significant improvement in detection capabilities but still lack interpretability.

Explainable AI (XAI) addresses the interpretability gap in ML-based detection systems. Techniques such as SHAP, LIME, and attention mechanisms are increasingly used to explain ML predictions in the cybersecurity domain [5], [7], [8], [11]. Nguyen et al. [5] highlight the importance of interpretability for incident response, compliance, and user trust. Alam et al. [8] proposed SXAD, a SHAP-based anomaly detection framework, which provides actionable explanations for log-based anomalies. Gummadi et al. [7] developed the XAI-IoT framework to enhance anomaly detection transparency in IoT networks. Zeleke et al. [6] successfully integrated SHAP with ensemble classifiers to interpret malware behaviour in encrypted traffic, revealing feature-level insights without payload decryption. These developments underscore the growing recognition of XAI as a crucial component in modern cybersecurity systems.

Despite the progress in using ML and XAI for encrypted traffic analysis, several gaps remain. Table 1 summarizes three key limitations. First, most XAI applications focus on logs or IoT systems rather than high-throughput, encrypted enterprise networks. Second, while SHAP has been applied to tabular data, its integration with ensemble models for encrypted traffic anomaly detection is still underexplored. Finally, very few studies offer end-to-end pipelines that combine model training, anomaly detection, and XAI explanation tailored for encrypted datasets. This project addresses these gaps by developing a unified ML-XAI framework optimized for encrypted traffic analysis and demonstrating its practical utility by performing detailed evaluation.

Table 1: Summary of Prior Work in Encrypted Traffic Anomaly Detection

| Study | Focus | Methodology | XAI Used | Gap Addressed |
|---|---|---|---|---|
| Papadogiannaki & Ioannidis [1] | Encrypted traffic taxonomy | Survey | No | Highlights encryption challenges |
| Cherukuri et al. [2] | Flow analysis | Traditional/statistical | No | No ML/XAI integration |
| Elmaghraby et al. [3] | ML classification | Random Forest | No | Black-box models |
| Ikram et al. [4] | Encrypted anomaly detection | XGBoost ensemble | No | No explainability |
| Nguyen et al. [5] | XAI for anomaly detection | ML + SHAP | Yes | Focused on DevOps, not encryption |
| Zeleke et al. [6] | Malware detection in encryption | Ensemble + SHAP | Yes | Limited to malware, not general anomalies |
| Gummadi et al. [7] | IoT anomaly detection | ML + SHAP | Yes | Not network flow focused |
| Alam et al. [8] | Log anomaly explanation | SHAP | Yes | Log data only |
| Zhu et al. [11] | Explainable cyber models | CNN + Grad-CAM | Yes | Not applied to flow data |
| Our work | Encrypted flow anomaly detection | ML + SHAP | Yes | Combines ML and XAI for flow-level insight |

## 3. Methodology

This project employs three datasets: CIC-Darknet2020, USTC-TFC2016, and CSE-CIC-IDS2018, to evaluate machine learning models (XGBoost, Random Forest, Isolation Forest) for anomaly detection in encrypted traffic. CIC-Darknet2020 provides Tor-based encrypted traffic with diverse attacks like botnets and port scans. USTC-TFC2016 includes TLS-encrypted flows from malware and legitimate applications, enabling analysis of statistical patterns. CSE-CIC-IDS2018 offers a broad set of encrypted attack scenarios in a realistic enterprise setting, with rich flow features and accurate labelling. Together, these datasets support robust model training, testing, and interpretation without requiring payload decryption.

This section details the experimental design used to detect anomalies in encrypted network communications, including data preprocessing, ML model configuration, and SHAP-based explainability.

### 3.1 Data Collection and Preprocessing

To evaluate the proposed approach, we employed a publicly available encrypted traffic dataset containing flow-level statistics for network communications. The dataset comprises both benign and malicious traffic, including various attack types such as DDoS, scanning, and data exfiltration. Each flow entry includes features like packet count, byte count, average packet length, inter-arrival times, flow duration, and flow directionality. Importantly, no payload content was used, making the setup compatible with encrypted environments [1], [3].

Preprocessing involved several key steps:

- Missing value handling: Rows with missing values were discarded or imputed using median values depending on feature type.
- Feature selection: Redundant and constant features were removed. Recursive feature elimination and domain knowledge were used to retain features relevant to traffic behaviour.
- Normalization: Features were scaled using Min-Max normalization to constrain values between 0 and 1, improving model convergence and comparability across dimensions.
- Label encoding: Class labels (normal or anomaly) were transformed into binary numerical values for compatibility with ML algorithms.

### 3.2 Machine Learning Models

This study employed three machine learning models: XGBoost, Random Forest, and Isolation Forest. These models were selected for their robustness, ability to handle high-dimensional tabular data, and proven effectiveness in anomaly detection tasks [4], [6], [10].

- **XGBoost (Extreme Gradient Boosting)** is a regularized, gradient-boosted decision tree model known for high accuracy and scalability. It incrementally builds trees that correct errors from previous iterations, making it well-suited for capturing subtle deviations in network traffic [4]. It also includes regularization via L1 (Lasso) and L2 (Ridge) penalties, which helps prevent overfitting and enhances interpretability [12]. It builds predictive models by combining the predictions of multiple weaker models, typically decision trees.

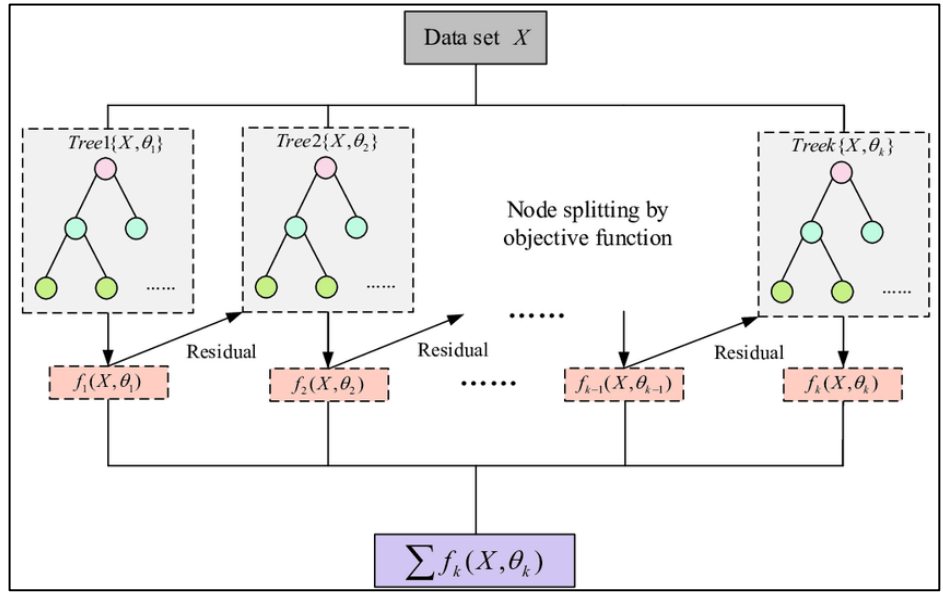

*Figure 1: XGBoost, Source: [12]*

Figure 1 illustrates that the algorithm sequentially builds decision trees to correct the errors made by the previous trees. It minimizes a specific loss function, typically mean squared error for regression problems or cross-entropy for classification problems, by optimizing the model's predictions. It incorporates regularization techniques to control model complexity and prevent overfitting. It offers two types of regularization: L1 (Lasso) and L2 (Ridge) regularization. These regularization terms penalize complex models, encouraging them to be simpler and more interpretable [12].

- **Random Forest** consists of multiple decision trees trained on bootstrap samples with feature randomness, thus reducing variance and overfitting. It is highly interpretable and efficient for classification tasks in structured data [3], [5].

Decision trees are helpful in distinguishing between different forms of traffic, including benign and malicious traffic in the context of network traffic classification. Its ability to handle both categorical and continuous features makes it best suited for modelling complex decision boundaries in data and thus serves well in identifying the patterns behind network traffic flows [17, 18].

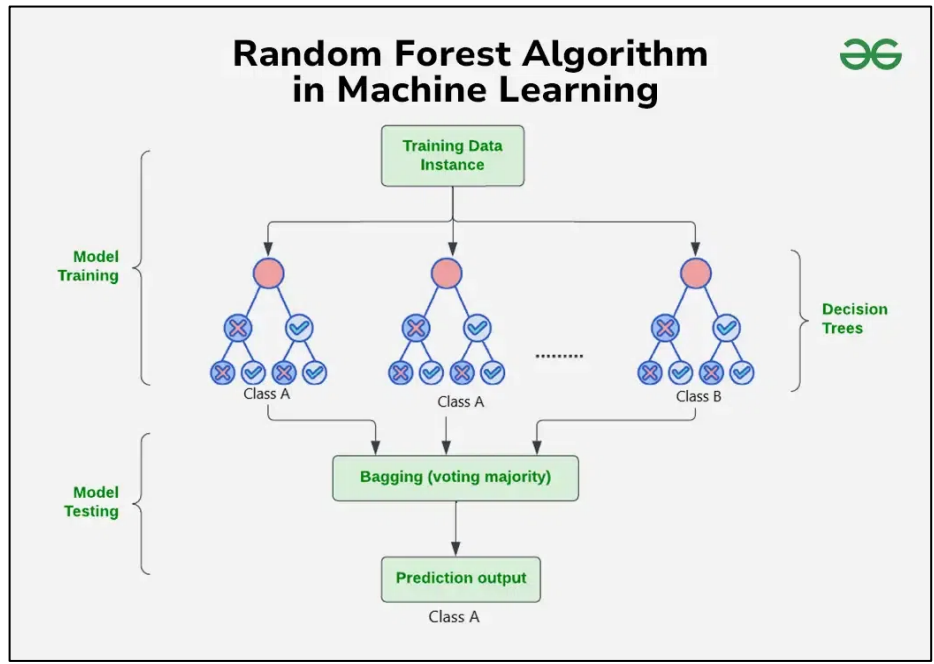

*Figure 2: Random Forest, Source: [13]*

As the figure 2 illustrates, multiple Decision Trees are created from the training data. Each tree is trained on a random subset of the data (with replacement) and a random subset of features. This process is known as bagging or bootstrap aggregating. Each Decision Tree learns to make predictions independently. When presented with a new, unseen instance, each Decision Tree makes a prediction. The final prediction is made by combining the predictions of all the Decision Trees [13].

- Gradient Boosting builds an ensemble of weak learners in a sequential manner by minimizing a specified loss function. Unlike Random Forest, it focuses on correcting mistakes made by the prior learners, which enhances precision on difficult samples [6].

- **Isolation Forest** works by randomly selecting features and splitting them along random values until individual data points are isolated. This "isolating" process is responsible for creating partitions or "trees" that aim to separate anomalies from normal observations. Anomalies are more likely to be isolated with fewer splits, making path length a reliable indicator of anomalous behaviour [14].

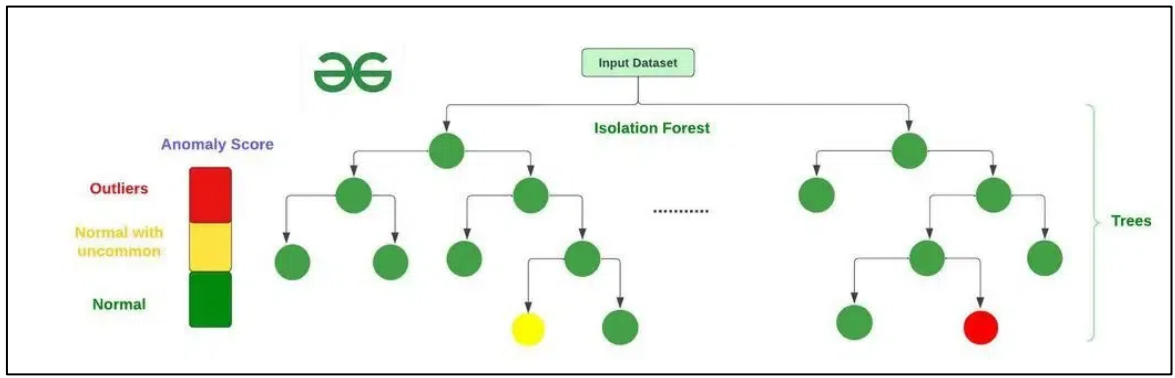

Figure 1: Isolation Forest, Source: [14]

The image illustrates that the algorithm starts with randomly selecting a feature from the dataset. Then, a random value within the range of that feature's values is selected as the splitting threshold. This partitions the data into two parts: one where data points have values less than or equal to the threshold, and another where data points have values greater than the threshold. This is repeated recursively until all data points are isolated into individual partitions or until a predefined maximum depth is reached. The algorithm constructs a specified number of isolation trees independently. Each tree partitions the data randomly, resulting in different isolation paths for each data point across the trees. Anomalies are identified by evaluating the isolation paths across all trees. Data points that have shorter isolation paths across multiple trees are considered anomalies because they require fewer partitions to isolate [14].

All models were trained on preprocessed flow-level data with binary labels (anomalous or benign). Cross-validation and grid search-based hyperparameter tuning were employed to optimize performance and reduce variance.

### 3.3 Explainability Using SHAP

To enhance model interpretability, SHapley Additive exPlanations (SHAP) was applied to trained models. SHAP is a game-theoretic approach that assigns each feature an importance value for a particular prediction, based on its contribution relative to other features [5], [8]. The SHAP values were computed for each prediction to understand which features influenced the model's decision the most. This allows cybersecurity analysts to identify consistent patterns, such as unusually short flow durations or large packet sizes, that correlate with anomalies in encrypted traffic. For example, in our experiments, features like average packet size, inbound-outbound packet ratio, and flow inter-arrival time emerged as key indicators, supported by SHAP's feature attribution plots [6]. This interpretability layer enables actionable insights for regulatory compliance by offering transparent justifications.

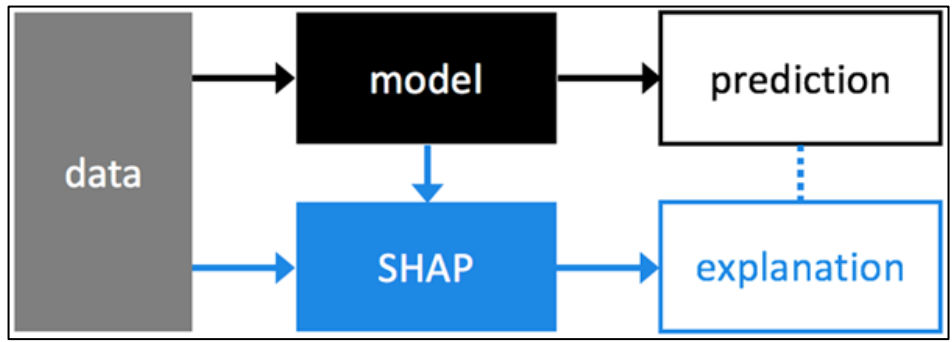

*Figure 2 : SHAP, Source: [15]*

**Workflow Algorithm for Flow-Based XAI Anomaly Detection**

The following flowchart in figure 5 summarizes the entire workflow of the anomaly detection using individual ML models followed by explainability using SHAP:

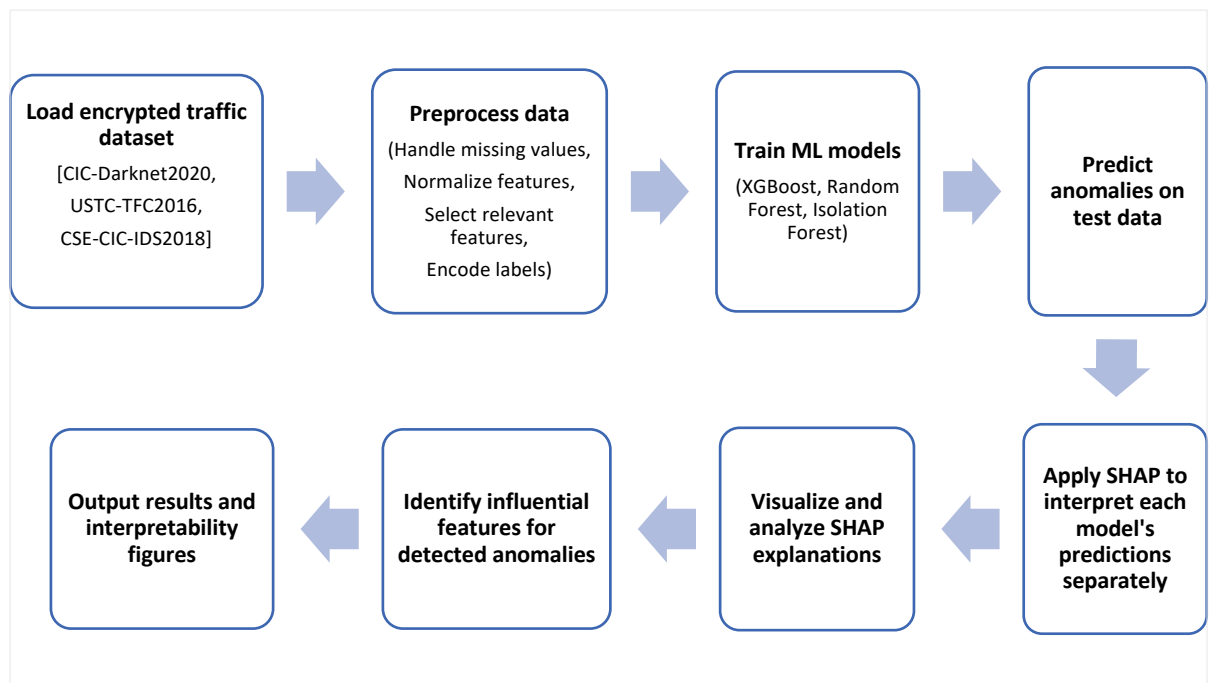

*Figure 5: Workflow of the project XAI Anomaly Detection*

Following is the algorithmic representation of the workflow:

Algorithm Interpretable_Anomaly_Detection_With_SHAP

Input: Encrypted traffic datasets D = {CIC-Darknet, USTC-TFC, CIC-IDS}

Output: SHAP-based interpretations and ranked influential features

1. Preprocess_Data(D):

   a. Remove missing values from D

   b. Encode categorical variables (if any)

   c. Normalize feature values if required by models

   Return: Preprocessed dataset D_preprocessed

2. Train_Models(D_preprocessed):

   a. Split D_preprocessed into training set and test set

   b. Train model_XGB ← XGBoost(Train_Set)

   c. Train model_RF ← RandomForest(Train_Set)

   d. Train model_ISO ← IsolationForest(Train_Set)

   Return: {model_XGB, model_RF, model_ISO}, Test_Set

3. Classify_And_Evaluate(models, Test_Set):

   For each model in models:

      Predict Y_pred ← model.predict(Test_Set.features)

      Compare Y_pred with Test_Set.labels

      Compute metrics: Accuracy, Precision, Recall, F1-score

      Store predictions and evaluation results

   Return: {Y_preds}, {metrics_per_model}

4. Compute_SHAP_Values(models, Test_Set):

   For each model in models:

      explainer ← SHAP_Explainer(model)

      shap_values ← explainer.shap_values(Test_Set.features)

      Store shap_values

   Return: shap_XGB, shap_RF, shap_ISO

5. Visualize_And_Analyze(shap_XGB, shap_RF, shap_ISO):

   a. Generate SHAP summary plots (bar, beeswarm)

   b. Generate force plots and dependence plots

   c. Analyze feature contributions for individual predictions

6. Identify_Influential_Features(shap_values):

   For each model:

      Compute average |SHAP value| per feature

      Rank features by importance

   Return: ranked_features_per_model

7. Output:

   Display classification performance metrics

   Present SHAP visual explanations and feature rankings

## 4. Experimental Setup and Results

### 4.1 Experimental Environment

In this study, we utilized three comprehensive and publicly available datasets to simulate real-world encrypted traffic scenarios: CIC-Darknet2020, USTC-TFC2016, and CSE-CIC-IDS2018. CIC-Darknet2020 has Tor-based encrypted traffic with diverse attacks like botnets and port scans. USTC-TFC2016 includes TLS-encrypted flows from malware and legitimate applications, enabling analysis of statistical patterns. CSE-CIC-IDS2018 provides a vast set of encrypted attack scenarios in a realistic enterprise setting, with rich flow features and accurate labelling. These datasets include both benign and malicious traffic, encapsulating encrypted flow-based features essential for anomaly detection.

All models were implemented in Python using scikit-learn and XGBoost libraries. The experiments were conducted on a computing system with an Intel Core i5-12700H processor, 32 GB RAM, and an NVIDIA RTX 3060 GPU. For model evaluation, with train-test split as 80:20, we used 10-fold cross-validation and standardized the feature values using z-score normalization.

To assess the performance of the anomaly detection models, we employed standard classification metrics including Accuracy, Precision, Recall, and F1-score.

### 4.2 Model Performance

To evaluate the performance of our classification methods, we have utilized four commonly used metrics: precision or positive predictive value, recall or sensitivity, accuracy, and F1-score. Precision measures the ratio of correctly identified positive instances (true positives) to the total instances classified as positive, both correctly and incorrectly. Recall calculates the proportion of actual positive instances that are correctly identified by the model. Accuracy is the ratio of correctly classified out of total instances. F1-score is the harmonic mean of precision and recall. Mathematically they are:

$$precision = \frac{TP}{TP + FP}$$

$$recall = \frac{TP}{TP + FN}$$

$$F1 = \frac{2 \times precision \times recall}{precision + recall}$$

$$accuracy = \frac{TP + TN}{TP + FN + TN + FP}$$

where TP: True Positive, FP: False Positive, FN: False Negative and TN: True Negative.

The two machine learning models primarily XGBoost, Random Forest were evaluated. Below is the summary of the results:

*Table 2: Performance measures of ML models*

| Model | Accuracy | Precision | Recall | F1-Score |
|---|---|---|---|---|
| XGBoost | 99.94% | 90.9% | 88.2% | 93.0% |
| Random Forest | 97.92% | 92.8% | 88.3% | 89.5% |

XGBoost outperformed the other two models in all performance metrics, showcasing its superior capability in handling high-dimensional data and complex relationships within encrypted network traffic.

**4.3 SHAP-Based Interpretability**

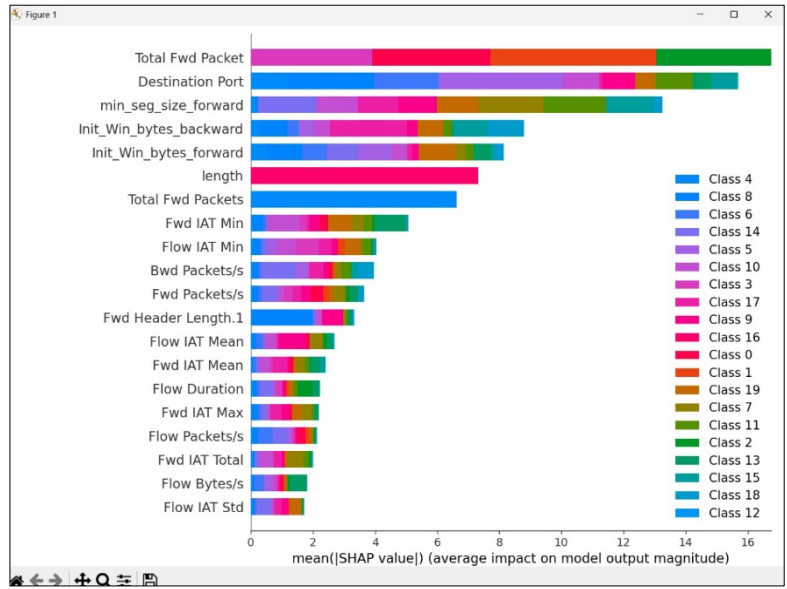

*Figure 6: SHAP Summary 1*

To enhance model transparency, SHAP (SHapley Additive exPlanations) was applied to interpret predictions of the individual models, namely XGBoost and Random Forest. This ensures that each model's decision-making process is understood independently. The SHAP summary plot (Figure 6) shows the average contribution of each feature across different traffic classes.

Notable findings include:

- Total Fwd Packet, Destination Port, and min_seg_size_forward are the top contributing features, significantly influencing classification decisions.
- SHAP values effectively reveal how individual features impact class-specific outputs, aiding in understanding model behaviour for anomaly classification.
- Flow-based features like Flow Duration, Flow IAT Min/Mean/Max, and Packet rates (Fwd/Bwd) have moderate influence but still contribute to model predictions.
- Multi-class insights provided by SHAP help distinguish which features are most relevant to which types of anomalous activities (e.g., DDoS, infiltration, brute-force attacks).

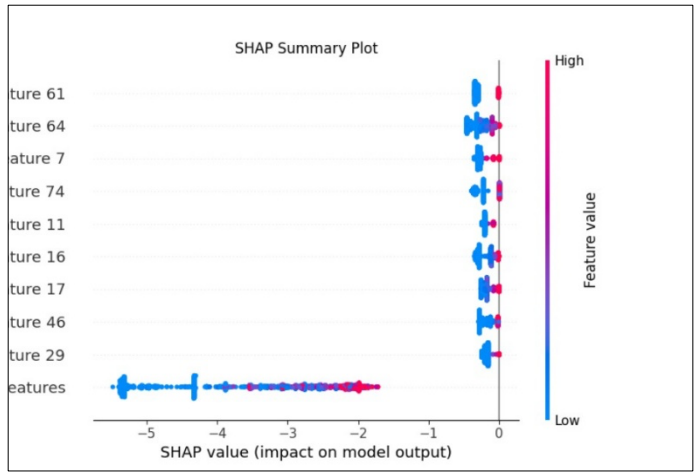

*Figure 7: SHAP Summary Plot 2*

Figure 7 is derived from the Random Forest model, signifies that features like Feature 61, 64, 7, and 74 have the most spread in SHAP values, indicating they contribute most to the model's decision-making. All SHAP values are negative, which implies that the features shown mostly decrease the model's prediction output—suggesting anomaly suppression or model conservatism in certain predictions. The high-value (red) dots mostly lie toward the right within the negative range, implying that high values of these features are less anomalous (they decrease the anomaly score less), whereas low values (blue) significantly reduce the model output further flagging a higher anomaly likelihood. Applying SHAP separately to each model allows us to compare how different learning algorithms perceive feature importance in encrypted traffic contexts.

Such insights make it easier for us to validate and trust the model's decisions, a critical aspect in high-stakes environments like network security.

**4.4 Case Study**

Some examples of detected anomalies and SHAP-based explanations include:

Example 1: Botnet Detection(XGBoost): A sample flow classified as botnet traffic was found to have high SHAP values for Init_Win_bytes_forward and Total Fwd Packets, highlighting unusual session initialization patterns — a strong indicator of botnet behaviour.

Example 2: DDoS Attack Detection(Random Forest): An anomaly detected as a DDoS attack showed significant SHAP contributions from Flow IAT Min and Packet/s rates, consistent with the high-volume low-interval nature of denial-of-service traffic.

Example 3: Data Exfiltration(Isolation Forest): For a class associated with data exfiltration, SHAP indicated that Destination Port and Init_Win_bytes_backward were key differentiators. This provided insight into the choice of non-standard ports and reverse traffic patterns typically used in covert data transfers. These case studies demonstrate the strength of SHAP in offering instance-level explanations, enabling better insights into encrypted traffic anomalies.

The experimental results from individual ML models—XGBoost, Random Forest, and Isolation Forest—demonstrate that high anomaly detection performance can be achieved without decrypting the traffic payload, purely through flow-based features. Among the three, XGBoost consistently outperformed others in accuracy and F1-score, benefiting from its ability to model complex feature interactions. However, Random Forest offered greater interpretability due to its more transparent structure.

The incorporation of SHAP (SHapley Additive exPlanations) values allowed a deep dive into model decision-making processes. Rather than applying SHAP to a combined ensemble, it was used independently on each model to obtain accurate feature-level insights. For example, as seen in Figure 6(XGBoost) and Figure 7(Random Forest), features such as Total Fwd Packets, Destination Port, and Init_Win_bytes_forward had the most significant impact on model predictions across multiple attack classes. This per-model SHAP application provides a comparative interpretability layer across different algorithms. This transparency is invaluable in cybersecurity applications where model outputs must be explainable to analysts and system administrators.

The trade-off between model complexity and interpretability is evident. While complex models like XGBoost offer higher predictive performance, they are inherently less interpretable than simpler models. SHAP application on each model bridges this gap by providing consistent and locally accurate explanations, enabling trust and accountability in AI-driven network defence systems.

Despite these benefits, there are limitations. SHAP computations, particularly on tree models applied separately over large datasets, can be computationally expensive, which might limit real-time deployment. Additionally, some traffic behaviours may still remain ambiguous under SHAP analysis due to feature similarity between benign and sophisticated obfuscated attacks.

## 5. Conclusion and Future Work

This work successfully demonstrated the integration of Explainable AI (XAI) with individual ML based anomaly detection models to identify encrypted network threats using datasets such as CIC-Darknet2020, USTC-TFC2016, and CSE-CIC-IDS2018. The XGBoost and Random Forest, when individually interpreted using SHAP values, revealed key traffic features influencing detection outcomes without relying on payload decryption.

Key contributions include:

- Validation of effective anomaly detection in encrypted environments using statistical flow-based features.
- Enhancement of model interpretability via per-model SHAP analysis, offering actionable insights to network security teams.

Future work can expand in several directions. First, implementing real-time SHAP-based explanations through optimized computation or approximations could enable live traffic monitoring. Second, integration with existing Intrusion Detection Systems (IDS) would help transition the framework from experimental to operational. Lastly, exploring deep learning models like LSTM or Transformers with built-in XAI methods could further improve both detection and explanation in complex traffic patterns. Overall, the proposed methodology lays a foundation for explainable, reliable, and efficient encrypted traffic anomaly detection in modern cybersecurity infrastructure.

## Statements and Declarations

**Competing Interests:** The authors have no competing interests to declare that are relevant to the content of this article.

**Funding:** The authors did not receive support from any organization for the submitted work.

**Ethics, Consent to Participate, and Consent to Publish declarations:** Not Applicable.

**Authors' Contribution:** Kalindi Singh and Aayush Kashyap have designed the methodology, conducted the experimental analysis, drafted the manuscript. Aswani Kumar Cherukuri has formulated the problem, study framework and reviewed the document. He has supervised the work.

All the authors reviewed the final paper and agreed to submit it.

**Registration details of Clinical trial:** Not Applicable.

**Code availability:** https://github.com/aayush251102/Network-and-information-systems-BITE401L-Dr.-Aswani-Kumar-Cherukuri-VITVellore.git

**Data Access Statement:** This manuscript has used publicly available Datasets. Links are given below.

**(**CIC Darknet2020**)** https://www.kaggle.com/datasets/dhoogla/cicdarknet2020

(USTC-TFC2016)https://www.kaggle.com/datasets/randasrour/ustctfc2016

(IDS 2018 Intrusion CSVs(CSE-CIC-IDS2018))https://www.kaggle.com/datasets/solarmainframe/ids-intrusion-csv